\documentclass{pasa}
\usepackage{import}
\usepackage{graphicx}
\usepackage{orcidlink}
\usepackage{soul}
\usepackage{longtable}
\usepackage{geometry}
\usepackage{appendix}
\usepackage{pdflscape}
\usepackage{lscape}
\usepackage{tablefootnote}

\title[The Parkes Observatory Pulsar Data Archive]{The Parkes Observatory Pulsar Data Archive}

\author[Toomey, L. J.]{Toomey, L. J.\orcidlink{https://orcid.org/0000-0003-3186-3266}$^{1}$,
Hobbs, G.\orcidlink{https://orcid.org/0000-0003-1502-100X}$^{1}$,
Dempsey, J.\orcidlink{https://orcid.org/0000-0002-4899-4169}$^{2,4}$,
Majewski, S.\orcidlink{https://orcid.org/0009-0004-2374-291X}$^{3}$,
Dai, S.\orcidlink{https://orcid.org/0000-0002-9618-2499}$^{1}$,
Reynolds, J. E.\orcidlink{https://orcid.org/0000-0001-9493-0935}$^{1}$
\affil{$^{1}$CSIRO Space \& Astronomy, P.O. Box 76, Epping, NSW 1710, Australia}
\affil{$^{2}$CSIRO Information Management \& Technology (IM\&T), PO Box 225, Dickson ACT 2602, Australia}
\affil{$^{3}$CSIRO Information Management \& Technology (IM\&T), Hobart TAS 7000, Australia}
\affil{$^{4}$Research School of Astronomy and Astrophysics, The Australian National University, Canberra, ACT 2611, Australia}
}

\jid{PASA}
\doi{10.1017/pas.\the\year.xxx}
\jyear{\the\year}

\usepackage{aas_macros}
\usepackage{hyperref} 
\hypersetup{colorlinks,citecolor=blue,linkcolor=blue,urlcolor=blue}

\begin{document}

\begin{frontmatter}
\maketitle

\begin{abstract}

Data from observations of pulsars made by Murriyang, the CSIRO Parkes 64-metre radio-telescope over the last three decades are more accessible than ever before, largely due to their storage in expansive long-term archives. Containing nearly 2 million files from more than 400 Parkes pulsar projects, CSIRO's Data Access Portal is leading the global effort in making pulsar data accessible. In this article, we present the current status of the archive, and provide information about the acquisition, analysis, reduction, visualisation, preservation and dissemination of these data sets. We highlight the importance of such an archive, and present a selection of new results emanating from archival data. 

\end{abstract}

\begin{keywords}
Parkes, pulsars, data, Data Access Portal, archive
\end{keywords}

\end{frontmatter}



\section{Introduction}

Observations of pulsars spanning from the 1990's to the present day by Murriyang\footnote{In the Wiradjuri Dreaming, Biyaami (Baiame) is a prominent creator spirit and is represented in the sky by the stars which also portray the Orion constellation. Murriyang represents the `Skyworld' where Biyaami lives.}, the CSIRO Parkes 64-metre radio-telescope, are archived for long-term storage in the CSIRO's Data Access Portal\footnote{\url{https://data.csiro.au}} (DAP), located in Canberra, Australia, and managed by CSIRO's Information Management and Technology (IM\&T) service. The archive is a historical record of snapshots of the sky as observed by Murriyang at frequencies in the radio band ranging from 0.4 to 24GHz, over a time span of 34 years.

\begin{figure*}
\centering
\includegraphics[width=400px]{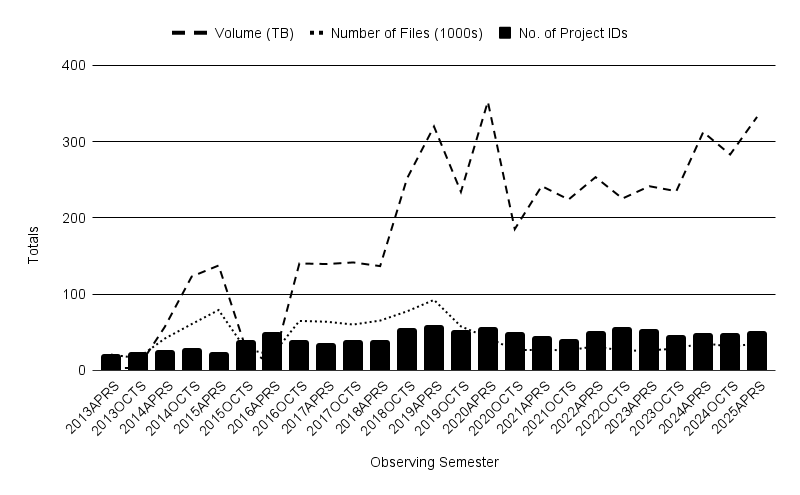}
\caption{Murriyang pulsar data published in CSIRO's Data Access Portal, by observing semester.}
\label{fg:dapPublishing}
\end{figure*}

At the time of writing, over 4.5 Petabytes of data from 450 unique Parkes pulsar-specific observation proposals (known as project identifiers, hereafter `PIDs') are publicly available for immediate download (Table \ref{tb:dataOverview}). Pulsar data ingested into DAP continues to grow at a steady rate (Figure \ref{fg:dapPublishing}), as Murriyang regularly enjoys upgrades of cutting-edge receiver technology.

\begin{table*}
  \caption{\textit{CSIRO's Data Access Portal - an overview of the data in the archive available for download at the time of writing.}}
  \begin{center}
    \begin{tabular}{p{0.5\textwidth}p{0.2\textwidth}}
      \hline
        Total number of Parkes project identifiers (PIDs)       & 450 \\
Total number of published pulsar collections 		    & 7014 \\
Range of observation dates                   		    & 1991\textendash{}2025 \\
Total data volume archived (Terabytes, TB)       	    & 4624 \\
Total number of published files              		    & 1796931 \\
\hline

    \end{tabular}
    \end{center}
  \label{tb:dataOverview}
\end{table*}

So why collect pulsar data and what makes Murriyang so important as an instrument? While pulsars are of considerable astrophysical interest in their own right, they are also important astrophysical tools --- searched for and subsequently monitored --- and observations have been used to understand many aspects of the known universe, for example stellar evolution (\citealt{2023MNRAS.523.5064C}), solar system dynamics (\citealt{2018MNRAS.481.5501C}), theories of gravity in strong field regimes (\citealt{2006Sci...314...97K}) and detection of a stochastic background of gravitational waves from pulsar timing experiments (\citealt{2023ApJ...951L...7R}).

For decades, the role of Murriyang in the discovery of new pulsars and the continuous stream of scientific results originating from pulsar observations has been remarkable. Murriyang has discovered 1235 out of the total number of 3748 known pulsars in the ATNF Pulsar Catalogue v2.6.3\footnote{\url{https://www.atnf.csiro.au/research/pulsar/psrcat/}}. Some of the highlights include the discovery of the only known double pulsar (\citealt{2003Natur.426..531B}), and the so-called `diamond planet' (\citealt{2011Sci...333.1717B}). Importantly for the purpose of this paper, archival data have also lead to a number of unexpected results --- for example the discovery of `Rotating Radio Transients' (RRATS, \citealt{McLaughlin_2006}) was made during re-processing of the Parkes Multibeam Pulsar Survey (PMPS, P268, \citealt{2001MNRAS.328...17M}), and the first unusually strong short-lived radio burst (a new class of object later dubbed `Fast Radio Bursts', FRBs) was discovered during re-processing of a survey of the Small Magellanic Cloud (\citealt{2007Sci...318..777L,2006ApJ...649..235M}).

Observations have also been carried out to study transient objects in the radio sky such as flare stars (\citealt{csiro:P1119-2021OCTS_04}), and Long-Period Transients (LPTs, \citealt{2023Natur.619..487H}). Spectral line and continuum observations are also supported - data from these observations are archived in the Australia Telescope Online Archive (ATOA\footnote{\url{https://atoa.atnf.csiro.au}}) and eventually migrated to the CSIRO ASKAP Science Data Archive (CASDA\footnote{\url{https://data.csiro.au/domain/casda}}). A recent addition to the suite of supported observing modes is the phase-resolved spectra mode, using the periodic on-off of known pulsars to study the emission and absorption spectra along the line of sight (\citealt{2025ApJS..278...13L}). Murriyang is also part of the Long Baseline Array network, supporting Very Long Baseline Interferometry (VLBI) observations including measurements of pulsar distances by parallax, e.g. \citealt{Dodson_2003}. Occasionally Murriyang is also used for confirmation and follow-up of point sources of interest, for example \citealt{2025ApJ...982L..53W}, an LPT discovered recently in data from the ASKAP radio telescope (\citealt{Hotan_Bunton_Chippendale_Whiting_Tuthill_Moss_McConnell_Amy_Huynh_Allison_et_al._2021}).

Archived data are from proposals ranging from targeted observations, for example of the globular cluster 47 Tucanae (PID P1006, \citealt{2019ApJ...885L..37Z}) to long-term monitoring programs like the Parkes Pulsar Timing Array (P456, \citealt{2013PASA...30...17M}), and Young Pulsar Timing (P574, \citealt{2008MNRAS.391.1210W}), to large sky surveys such as the PMPS and SUPERB - A SUrvey for Pulsars \& Extragalactic Radio Bursts (P858, \citealt{2018MNRAS.473..116K}). The DAP also contains data sets that relate to a particular publication, software package or data release.

Data in the DAP are embargoed for a period of 18 months before being released for public use in `collections' grouped by observation semester (nominally with two semesters annually). Embargoed data are only accessible to Principal Investigators (PIs) and contributors to a proposal. All collections are labeled with a unique Digital Object Identifier\footnote{\url{https://www.doi.org/}} (DOI) that is persistent with the life of the collection, thereby providing a mechanism to couple scientific research with good provenance.

The importance of such an archive cannot be underestimated, and it continues to yield new results when the data are run through new algorithms --- here are just a few examples. Reprocessing of the Parkes Multibeam survey with a GPU-accelerated processing pipeline recently yielded 37 new pulsars (\citealt{2023MNRAS.522.1071S}). Artificial Intelligence and Machine Learning tools are also playing a more significant role in candidate and/or anomaly detection (\citealt{10.1093/mnras/staf046}). Recently a search of archival data was conducted to question the repeating nature of some transient events (\citealt{2024ApJ...974..248Z}), and in another example, the authors mined archival observations of Open Clusters for potential candidate pulsars (\citealt{2025arXiv250619236Z}). 16 years after the discovery of the first FRB, an additional one was found in the same data-set (\citealt{10.1093/mnrasl/slz023}).

This paper is intended as a follow up to 
 \citealt{2011PASA...28..202H}, bringing users up to date with major developments in the archive since 2011, and describing how the archive plays an important role as we follow the data on a journey from the telescope to the end user, with steps in place to ensure that data quality remains at a high standard throughout.

In Section 2 we describe aspects of data acquisition, including the importance of pulsar data for the field of radio astronomy, observation types, data formats and archive provision. Section 3 focuses on data preservation, including archive structure and scope of available data products. Aspects of data dissemination are described in Section 4, and in Section 5 we introduce the \textsc{PFITS} software package and briefly discuss data reduction methods and visualisation. In Section 6 we discuss the challenges and future requirements for pulsar data archiving in the era of accelerated data volume acquisition, and leveraging Cloud platforms for processing of DAP data. 

\section{Data Acquisition}

The pulsar data products from Murriyang can be thought of as a snapshot of the sky at a particular time and radio frequency, and are generally either termed `fold-mode', `search-mode', or `calibration-mode', depending on the observation type. These data products are described in paper I (\citealt{2011PASA...28..202H}), but introduced briefly here.

Fold-mode observations are from a pointing of a known pulsar, where the data are `folded' or stacked at the known rotation period of the pulsar, to form an integrated pulse profile that is averaged over a time period longer than the pulsar's spin period --- in the DAP, files of this type have the extension `.rf'.

Prior to 2018, these fold-mode files were also averaged over all frequency channels, all polarisations and integrations, to create a separate sub-set of files\footnote{An exception to this applies to data from the PULSE@Parkes outreach project (P595, \url{https://research.csiro.au/pulseatparkes}), where the frequency resolution was averaged to 8 channels} --- an accompanying preview image of the integrated pulse profile is also available allowing the user to make a judgment on observation quality.

Search-mode observations comprise a multi-channel stream of data with time (a `time series') for the purpose of searching a particular sky location for radio signals, periodic or otherwise. Files of this type have the extension `.sf'. These observations make up 93\% of the total archive volume.

Two types of calibration-mode observations are used for pulsar observations taken with the current receiver suite. The first type is of a waveform injected into the signal path (\citealt{2020PASA...37...12H}), and the second is from observations of a reference radio source with a known stable flux density, notably Hydra A or more recently PKS B1934-638. The injected signal is used to calibrate the polarisation information of the astronomical data, and generally taken before (and sometimes after) a pulsar observation. The reference radio source provides a means of calibrating the flux density of an observation. The calibration-mode files have the extension `.cf'.

\subsection{Archival data products}

The accepted data format for pulsar data in the DAP is `PSRFITS' (\citealt{2004PASA...21..302H}), although archiving of other research data and/or software is also supported. PSRFITS is a flexible and extensible format based on the Flexible Image Transport System (FITS, \citealt{2010A&A...524A..42P}) specifically for pulsar data, adhering to the current version of the definition\footnote{\url{https://www.atnf.csiro.au/research/pulsar/psrfits_definition/Psrfits.html}}.

At the completion of a particular pulsar observation, the data are transferred to a staging server where they are checked for integrity, converted to the required format if required, and sorted into collections by PID and observation semester. Finally metadata and checksums are captured, and the collections are placed in the DAP upload queue. Once in the DAP staging area, checks to verify both integrity and metadata are applied, and if successful, the data then progress through to final publication where they become accessible via the DAP's web-based portal.

PSRFITS format was not always supported by instrumentation at Murriyang. For example the BPSR/HIPSR (\citealt{2016JAI.....541007P}) pulsar and spectral line data acquisition system (or `backend') produced search-mode files but in the native filterbank\footnote{\url{https://sigproc.sourceforge.net}} format. Collating and conversion of early archival data is an ongoing process --- data from the digital backends such as the Analogue Filterbank (\citealt{manchester2001parkes}) and BPSR continue to be converted to PSRFITS format on a dedicated virtual compute host in CSIRO's Bowen Research Cloud (BRC) prior to being published on DAP.

\section{Data Preservation}

To ensure preservation of our pulsar data products and to encourage future reuse, every file undergoes strict confirmation that they adhere to the PSRFITS definition, including check-summing and metadata completeness prior to archiving. 

\subsection{Accepted file formats} 

A PSRFITS format file consists of a primary Header Data Unit (HDU) containing observation metadata, followed by a series of binary extension HDUs, storing metadata and history specific to an observation, and the associated data products.

These files are readable by open-source software packages for pulsar data analysis such as \textsc{PSRCHIVE} (\citealt{2004PASA...21..302H}), \textsc{PRESTO} (\citealt{2011ascl.soft07017R}), the \textsc{PFITS} package (described in Section 5), and FITS file viewers such as NASA HEASARC \textsc{Fv}\footnote{\url{https://heasarc.gsfc.nasa.gov/docs/software/ftools/fv/}}.

The pulsar astronomy community generally sees the benefit of storing data in a format that allows for metadata updates and the ability to add entirely new HDUs if required. However, FITS cannot store data streams from receivers with multiple beams such as the recently commissioned Cryogenically-cooled 72-beam Phased Array Feed (CryoPAF) on Murriyang, and is not suited for appending large data-sets. We are currently trialling Spectral-Domain Hierarchical Data Format (SDHDF, \citealt{TOOMEY2024100804}) as a replacement file format.

\subsection{Data provenance}

The history of the origin, processing, and methodologies associated with a particular data-set encompass the provenance. For pulsar data products from Murriyang, the file metadata and DAP policies provide a high-level of provenance in a number of ways:
\begin{itemize}
  \item Comprehensive metadata capture in the PSRFITS headers, for example key dates, astronomical source information, receiver and digital acquisition system information.
  \item Use of a system of Digital Object Identifiers (DOIs) and persistent links to collections, an assurance is given to an author of a particular publication that a DAP collection of associated data (or for software, a specific version) will be accessible for the life of the archive.
  \item The transfer of large volumes of data over a network can occasionally lead to data loss --- this can be problematic if a user wishes to reproduce a set of published results --- the DAP ensures that checksums of individual files are stored in the metadata, thus providing the user the assurance that the data product is identical to when it was first archived.
  \item DAP provides the citation text for each collection --- this ensures that data are correctly cited in peer-reviewed publications.
\end{itemize}

Data provenance is crucial in order to reproduce scientific results --- in 2017, we attempted to re-create reprocessing of fold-mode pulsar data on virtual machines across multiple operating system environments\footnote{\url{https://doi.org/10.4225/08/56B4094AA4D01}}. The team found that by using containerised operating systems, and the DOIs provided by the DAP, they were able to fully reproduce the software and data environments and to reproduce the data perfectly. (They were only able to partially reproduce analysis results, but this was due to a random seed built in to the processing software.)

\subsection{DAP collection types}

The DAP groups data products in `collections'. The types of collections are grouped broadly as `standard pulsar' and `other pulsar-related' --- the latter can be research data and/or software.

\subsubsection{`Standard' pulsar collections}

Each observing semester, astronomers can submit a proposal for observing time with ATNF's telescopes through the ATNF OPAL\footnote{\url{https://opal.atnf.csiro.au}} system --- these include Non A-priori Assignable (NAPA) proposals that may over-ride allocated observing for rapid follow-up of a transient source for example. The proposals are judged on scientific merit and observing time is allocated accordingly --- these are referred to as `standard' pulsar collections and contain data from fold-, search-, or calibration-mode observations in PSRFITS format. These collections are bundled by semester, for example a P456 observation in April 2024 can be found in the 2024APR semester (2024APRS spans the 6 months from April 1st to September 30th 2024). A P456 observation in January 2025 will be bundled in the 2024OCT semester (2024OCTS spans from October 1st 2024 to March 31st 2025). The metadata for these `standard' collections (proposal team details, descriptions, embargoes and license) are generated automatically from the proposal in the OPAL system.

Observation time can be granted at short notice and without an OPAL proposal through applying for Director's Discretionary Time or Target of Opportunity (ToO) time. An example of this might be a follow-up of a new source --- in this case the Project ID assigned is prefixed with a `PX', and prior to 2025, were collated in DAP collections with the title `PUNDEF', for projects that are undefined in the OPAL system. There are three Project IDs that were exceptions to this rule however --- PX500 and PX501 were assigned to projects that had purchased telescope time, and PX600 was assigned to observations from the Breakthrough Listen\footnote{\url{https://seti.berkeley.edu/listen/}} initiative.

\subsubsection{`Other' pulsar-related collections}

The DAP also archives pulsar-related data-sets that are not necessarily observation data and do not fit into the `standard' type classified above. These collections may be data or software products related to a specific publication or project. Some examples of these are:
\begin{itemize}
\item The Parkes Pulsar Timing Array (PPTA) published their first, second and third data releases for general use.
\item Johnson \& Kerr et al. 2017 (\citealt{10.1093/mnras/stx3095}) published their polarimetry dataset, referenced from their publication by the DOI.
\item Software releases for the ATNF Pulsar Catalogue are published on a regular basis. 
\end{itemize}
These and `other' pulsar-related collections and the publications they are referenced in, including their persistent DOIs are shown in  Table~\ref{tab:dataOther}.

\begin{table*}
\small
  \caption{\textit{A selection of `Other' pulsar-related collections grouped by subject matter, their collection DOI, and where they are referenced.}}
  \begin{center}
  \small
  \begin{tabular}{p{0.5\textwidth}p{0.5\textwidth}}
   \hline
    Collection Title & Collection DOI \\
\hline
Pulsar Timing & \\
\hline
The Parkes Pulsar Timing Array (PPTA) Data Release 1                           & http://doi.org/10.4225/08/534CC21379C12 (\citealt{2013PASA...30...17M}) \\
The 23 year PSR B1259-63 dataset                                               & https://doi.org/10.4225/08/5318FF909B6DD (\citealt{10.1093/mnras/stt2123}) \\
PPTA pulsar data set from Reardon et al. (2015)                                & http://doi.org/10.4225/08/561EFD72D0409 (\citealt{10.1093/mnras/stv2395}) \\
Madison et al. data set for gravitational wave search                          & https://doi.org/10.4225/08/560A00E2036F6 (\citealt{10.1093/mnras/stv2534}) \\
PPTA pulse profiles                                                            & http://doi.org/10.4225/08/54F3990BDF3F1 (\citealt{10.1093/mnras/stv508}) \\
The Parkes pulsar calibration data release                                     & http://doi.org/10.4225/08/58363e853a58b \\
Pulsar Polarimetry at 1.4 GHz from Johnston and Kerr (2017)                    & http://doi.org/10.4225/08/59952c840ae35 (\citealt{Johnston_2017}) \\
Comparison of pulsar positions from timing and very long baseline astrometry   & http://doi.org/10.4225/08/58a0e1593c5be (\citealt{10.1093/mnras/stx837}) \\
Data set for Parkes Pulsar Timing Array constraints on ultralight scalar field dark matter & https://doi.org/10.25919/5bc67e4b7ddf2 (\citealt{Porayko_2018}) \\
Data set for modelling scintillation of PSR J1141-6545                         & https://doi.org/10.4225/08/5afe339d597f9 (\citealt{10.1093/mnras/stz643}) \\
Data files relating to "A pulsar-based timescale from the international pulsar timing array" & https://doi.org/10.25919/5c354f2623ac5 (\citealt{10.1093/mnras/stz3071}) \\
Parkes Pulsar Timing Array Data Release 2                                      & https://doi.org/10.25919/5db90a8bdeb59 (\citealt{Kerr_Reardon_Hobbs_Shannon_Manchester_Dai_Russell_Zhang_vanStraten_Osłowski_etal._2020}) \\
Timing analysis of the PPTA data release 2                                     & https://doi.org/10.25919/cx59-a798 (\citealt{10.1093/mnras/stab1990}) \\
Evaluating the prevalence of spurious correlations in simulated pulsar timing array datasets & https://doi.org/10.25919/3yj4-rx31 (\citealt{10.1093/mnras/stac2100}) \\
Parkes Pulsar Timing Array Third Data Release (part 1 of 2)                    & https://doi.org/10.25919/j4xr-wp05 (\citealt{2023PASA...40...49Z}) \\
Parkes Pulsar Timing Array Third Data Release (part 2 of 2)                    & https://doi.org/10.25919/axvw-qa43 (\citealt{2023PASA...40...49Z}) \\
Parkes Pulsar Timing Array data sets for PSR J0437-4715                        & https://doi.org/10.25919/20rx-5f63 (\citealt{Reardon_2024}) \\
\hline
Pulsar and Interstellar Medium properties & \\
\hline
Dynamic spectra for PSR J0437-4715                                             & https://doi.org/10.25919/5f3cd2bc1c213 (\citealt{Reardon_2020}) \\
Data files from the Parkes UWL receiver system for PSR J1803-3002A in NGC 6522 & https://doi.org/10.25919/5f45d801827d6 (\citealt{2020ApJ...905L...8Z}) \\
Flux density variability of 286 radio pulsars from a decade of monitoring      & https://doi.org/10.25919/14zq-a803 (\citealt{2021MNRAS.501.4490K}) \\
A polarization census of bright pulsars using the Ultra-Wideband Receiver on the Parkes radio telescope & https://doi.org/10.25919/gptm-d012 (\citealt{Sobey_2021}) \\
Dynamic Spectra for PSR J1603-7202                                             & https://doi.org/10.25919/82f5-mh79 (\citealt{Walker_2022}) \\
\hline
Single pulse and FRB properties & \\
\hline
Parkes observations of fast radio bursts FRB 171209, FRB 180309, FRB 180311 and FRB 180714 & https://doi.org/10.25919/5cb0344970ef3 (\citealt{10.1093/mnras/stz1751})  \\
Database of Single pulses from the Parkes telescope                            & https://doi.org/10.25919/5e33a52c18a17 (\citealt{Zhang_2020}) \\
\hline
Simulations & \\
\hline
SPARKESX: Single-dish PARKES data sets for finding the uneXpected - Part 1     & https://doi.org/10.25919/fd4f-0g20 (\citealt{10.1093/mnras/stac2558}) \\
SPARKESX: Single-dish PARKES data sets for finding the uneXpected - Part 2     & https://doi.org/10.25919/sqa9-rp38 (\citealt{10.1093/mnras/stac2558}) \\
SPARKESX: Single-dish PARKES data sets for finding the uneXpected - Part 3     & https://doi.org/10.25919/4g8p-gd74 (\citealt{10.1093/mnras/stac2558}) \\
\hline
Miscellaneous & \\
\hline
Interesting pulsar data used in the Amazon Cloud Prototype                     & https://doi.org/10.4225/08/56B4094AA4D01 \\
Vela pulsar data with the Parkes testbed facility                              & https://doi.org/10.4225/08/59183e949e033 (\citealt{Sarkissian_2017}) \\
PSRCAT v2: The ATNF Pulsar Catalogue                                           & https://doi.org/10.25919/tebw-ds72 (Hobbs et al. in prep.) \\
\hline

  \end{tabular}
  \end{center}
  \label{tab:dataOther}
\end{table*}

\subsection{Embargo overview}

Each PID has a PI and often multiple contributors. The embargoed files from a particular collection are accessible for the PI and contributors with approved credentials. Once the specified embargo period has lapsed, the files then become publicly accessible and available for download.

The proprietary period of a particular collection can be extended or removed if required. One example of this is the PULSE@Parkes project (P595, \citealt{2009PASA...26..468H}) --- an outreach program designed to involve high school students from around the world in real-time observations and pulsar data processing using Murriyang --- data from which are made publicly available immediately after the observations. Projects with paid time on the telescope may choose to extend the embargo beyond the proprietary period.

\subsection{Scope of available collections}

In this subsection, we present an overview of the available published collections in the DAP at the time of writing, including the scope of data by receiver and digital backend, sources and sky coverage. We also present a list of collections containing discoveries that provided breakthroughs in our understanding of pulsar classification and astrophysics.

\subsubsection{Scope of available receiver and data acquisition instrumentation}

The ability of Murriyang to continue to provide cutting-edge science in the field of radio astronomy is in part due to regular updates and replacements of the receivers and the digital acquisition systems. Recent additions are the Ultra-wide Bandwidth Low (UWL) receiver providing simultaneous bandwidth from 0.7 to 4 GHz (\citealt{2020PASA...37...12H}), and the CryoPAF, both developed in-house by the ATNF receiver group (paper in prep.). 

Table~\ref{tab:dataRx} is a comprehensive list of Murriyang's receiver fleet to date. This single historic record of the instrumentation since the early 1980's is included here to provide context for the reader --- not all receivers observed pulsars, and for those that did, not all have data that are accounted for. We are always on the lookout to publish historic archival data in the DAP --- by providing this comprehensive list, we hope that these missing data come to light. The AT Multi-Band receiver (see \url{https://www.atnf.csiro.au/observers/memos/d95b8a~1.pdf} and the Fourth Annual Report of the Australia Telescope Project (CSIRO, Oct. 1987, p.9, Appendix B)$^a$) was a five-feed receiver package --- the S/X-band$^1$ was a special concentric dual-band feed allowing S and X bands to be observed simultaneously, mostly for astrometric VLBI. The Methanol$^2$ (also known as the `Old Meth'), SETI$^3$, K/KU-band$^4$, Galileo$^5$, 10-50$^6$ and 13MM$^7$ (see \url{http://hdl.handle.net/102.100.100/109880?index=1}) receivers were all dual-feed packages. In 2000, the Methanol$^2$ `FRONTEND' parameter key was changed from `METHANOL' to `METH6' and `METH12' to reflect the two independent feeds of the receiver. The 50 cm frequency band of the 10-50 receiver was shifted upwards during it's time on the telescope in order to avoid phone and Digital TV interference --- from 2003 to July 2009, the range was 680 +/- 32 MHz, then 685 +/- 32 MHz, before finally settling on 700 to 764 MHz. Confusion about the polarisation of the 13MM dual-band receiver$^7$ is evident in the data --- the `13MM' parameter key was used for both feeds, causing uncertainty about whether the polarisation parameters were set correctly. From historical records, we have ascertained that the narrow-band receiver was used predominantly for VLBI and that all are circular if 21GHz < frequency < 23GHz. Our records also show that data prior to 21/11/2013 were marked as linear, but after were marked as circular, regardless of which feed was actually used. 

In some cases, data were recorded with parameter values that did not follow the traditional naming scheme --- for completeness these additional parameters found in the headers of data from some receivers are listed in Table~\ref{tab:dataRxMisc}.

\begin{landscape}
\begin{table}

\small
\caption{\textit{Murriyang's receiver fleet since the early 1980's --- used for both pulsar and non-pulsar observations. The `FRONTEND' field refers to the value of the `FRONTEND' parameter key in a PSRFITS file primary HDU (note, keys marked with $^!$ indicate that there are no PSRFITS files found in the DAP). The `Polarisation' field indicates the number and type of polarisation of the feed, linear (LIN) or circular (CIRC). Acronyms are as follows: Australia Telescope (AT), National Radio Astronomy Observatory (NRAO), Search for Extraterrestrial Intelligence (SETI), Dominion Radio Astrophysical Observatory (DRAO), Global Magneto-Ionic Medium Survey (GMIMS), Max Planck Institute (MPI). The contents of this table was created from Parkes schedule archives and an online receiver database$^b$ (from 1998 on-wards), and otherwise referenced in line where known.}}
\begin{tabular}
{p{0.33\textwidth}p{0.1\textwidth}p{0.1\textwidth}p{0.06\textwidth}p{0.11\textwidth}p{0.09\textwidth}p{0.38\textwidth}}
\hline
Name & FRONTEND & Frequency range (GHz) & No. of pixels & Polarisation (No., type) & Operational & Reference\\
\hline
K-band maser receiver                  & K-BAND$^!$           & 22\textendash{}24          & 1  & 2, CIRC     & 1982\textendash{}1994    & ATNF Annual Report 1994 \\
843MHz disc feed                       & 843MHZ$^!$           & 0.841\textendash{}0.845    & 1  & 2, LIN      & 1986\textendash{}1991    & \citealt{1993AJ....105.1666G} \\
Broad-band H-OH receiver               & H-OH                 & 1.2\textendash{}1.8        & 1  & 2, LIN/CIRC & 1993\textendash{}2016    & ATNF Annual Report 1993 \\
Q-band maser receiver                  & Q-band$^!$           & 42.4\textendash{}43.5      & 1  & 1, LIN      & 1986\textendash{}1995    & \citealt{Hall_Wark_Wright_1987} \\
AT Multi-Band receiver (L-band)        & L-BAND$^!$           & 1.3\textendash{}1.8        & 1  & 2, LIN      & 1985\textendash{}1997    & $^a$ \\
AT Multi-Band receiver (S-band)        & S-BAND$^!$           & 2\textendash{}2.5          & 1  & 2, LIN      & 1985\textendash{}2003    & $^a$ \\
AT Multi-Band receiver (C-band)        & C-BAND               & 4.5\textendash{}5          & 1  & 2, LIN/CIRC & 1985\textendash{}2020    & $^a$ \\
AT Multi-Band receiver (X-band)        & X-BAND$^!$           & 7.9\textendash{}9.3        & 1  & 2, LIN/CIRC & 1985\textendash{}2012    & $^a$ \\
AT Multi-Band receiver (S/X-band)$^1$  & S-band$^!$           & 2\textendash{}2.5          & 1  & 2, CIRC     & 1985\textendash{}2018    & $^a$ \\
AT Multi-Band receiver (S/X-band)$^1$  & X-band$^!$           & 7.9\textendash{}9.3        & 1  & 2, CIRC     & 1985\textendash{}2018    & $^a$ \\
Methanol dual-band receiver$^2$        & METH12$^!$           & 12\textendash{}12.8        & 1  & 2, LIN      & 1989\textendash{}2002    & ATNF Annual Report 1992 (p.26) \\
NRAO 7-beam Multibeam receiver         & NRAO$^!$             & 4.55\textendash{}5.15      & 7  & 2, LIN      & 1990                     & \citealt{1993AJ....105.1666G} \\
50cm cooled receiver                   & 50CM$^!$             & 0.6\textendash{}0.8        & 1  & 2, LIN      & 1990\textendash{}2003    & (Unknown) \\
70cm/75cm cavity-backed receiver       & 70CM	              & 0.420\textendash{}0.452    & 1  & 2, LIN      & 1991\textendash{}2003    & \citealt{1996MNRAS.279.1235M} \\ 
Methanol dual-band receiver$^2$        & METH6                & 6\textendash{}7            & 1  & 2, CIRC     & 1991\textendash{}2020    & ATNF Annual Report 1992 (p.26) \\ 
75MHz Erickson Feed                    & 75MHZ$^!$            & 0.044\textendash{}0.092    & 1  & 2, LIN      & 1993\textendash{}2014    & \citealt{1995ApJ...454..125E} \\
SETI dual-band receiver (L-band)$^3$   & L-band$^!$           & 0.995\textendash{}1.745    & 1  & 2, LIN      & 1995                     & ATNF Annual Report 1994 \\
SETI dual-band receiver (S-band)$^3$   & S-band$^!$           & 1.745\textendash{}3        & 1  & 2, LIN      & 1995                     & ATNF Annual Report 1994 \\
K/KU-band dual-band receiver$^4$       & KU-BAND$^!$          & 12\textendash{}18          & 1  & 2, LIN      & 1995\textendash{}2015    & ATNF Annual Report 1994 \\
K/KU-band dual-band receiver$^4$       & K-BAND$^!$           & 20\textendash{}25          & 1  & 2, LIN/CIRC & 1995\textendash{}2008    & ATNF Annual Report 1994 \\
Galileo receiver$^5$                   & GALILEO              & 2.27\textendash{}2.32      & 1  & 2, CIRC     & 1996\textendash{}present & \citealt{584502} \\
21cm Multibeam receiver                & MULTI                & 1.250\textendash{}1.550    & 13 & 2, LIN      & 1997\textendash{}2020    & \citealt{1996PASA...13..243S} \\  
10-50 dual-band receiver$^6$           & 1050CM               & 2.65\textendash{}3.55      & 1  & 2, LIN      & 2003\textendash{}2018    & \citealt{1532537} \\
10-50 dual-band receiver$^6$           & 1050CM               & 0.7\textendash{}0.764      & 1  & 2, LIN      & 2003\textendash{}2018    & \citealt{1532537} \\
Mars receiver                          & MARS                 & 7.9\textendash{}9.1        & 1  & 2, CIRC     & 2003\textendash{}present & ATNF Annual Report 2003 (p.61) \\ 
Galileo receiver (broad-band)$^5$      & GALILEO\_B           & 2\textendash{}2.5          & 1  & 2, CIRC     & 2004\textendash{}present & \citealt{584502} \\
Methanol Multibeam receiver            & METHMB               & 6\textendash{}6.7          & 7  & 2, CIRC     & 2006\textendash{}2010    & \citealt{2007IAUS..237..403C}$^*$  \\ 
DRAO/GMIMS feed                        & DRAO                 & 0.3\textendash{}0.9        & 1  & 2, LIN      & 2008\textendash{}2012    & \citealt{Wolleben_2019}$^*$  \\
13MM dual-band receiver$^7$            & 13MM                 & 16\textendash{}26          & 1  & 2, LIN      & 2008\textendash{}present & ATNF Annual Report 2010 (p.90) \\ 
13MM dual-band receiver$^7$            & 13MM                 & 21\textendash{}22.3        & 1  & 2, CIRC     & 2008\textendash{}present & ATNF Annual Report 2010 (p.90) \\ 
Bonn/MPI Phased Array Feed             & MPIPAF$^!$           & 0.7\textendash{}1.7        & 36 & 2, LIN      & 2016                     & \citealt{2016ceaa.conf..909C} \\
Ultra-wide Bandwidth Low receiver      & UWL	              & 0.704\textendash{}4.032    & 1  & 2, LIN      & 2018\textendash{}present & \citealt{2020PASA...37...12H} \\  
Cryogenically-cooled Phased Array Feed & CRYOPAF              & 0.7\textendash{}1.95       & 72 & 2, LIN      & 2025\textendash{}present & (in prep.) \\ 
\hline

\end{tabular}
\label{tab:dataRx}
\end{table}
\end{landscape}

\clearpage

\begin{table*}
\small
\caption{\textit{Other `FRONTEND' parameter values found in the PSRFITS primary HDU of some early pulsar data.}}
\begin{center}
\begin{tabular}{llll}
\hline
FRONTEND & Frequency range (GHz) & No. of beams & Notes \\
\hline
MULT\_1, 20cm\_MB\_1 & 1.250\textendash{}1.550 & 1 & Beam 1 of the 21cm Multibeam receiver \\ 
1010CM               & 2.6\textendash{}3.6     & 1 & 10cm feed of the 10-50 dual-band receiver \\ 
5010CM, 50CM         & 0.7\textendash{}0.764   & 1 & 50cm feed of the 10-50 dual-band receiver \\
\hline

\end{tabular}
\end{center}
\label{tab:dataRxMisc}
\end{table*}

The data acquisition instrumentation produces a binary data stream from the analogue sky signal that becomes the astronomy data products. Many collaborative efforts since the early 1990's have contributed to build, configure and update these systems on Murriyang -- Table~\ref{tab:dataBackend} lists these systems and the collaborations involved, and detailed specifications are presented in Table~\ref{tab:dataBackend2}.

\begin{table*}
\small
\caption{\textit{Murriyang's principal data acquisition systems used mainly for pulsar observations, believed to be complete since 1990 --- referenced where possible, showing the year they were commissioned, number of years in service (in brackets), and development credit. `INSTRUMENT' refers to the value of the `INSTRUMENT' parameter key in a PSRFITS file. The S2\textsuperscript{*} recorder was installed for VLBI observations but also used for pulsar observations from 1996 to 1999, and then returned to VLBI use only until 2002. BPSR\textsuperscript{**} was also known as the HI Parkes Swinburne Recorder, HIPSR. Apollo\textsuperscript{+} is a software instance running on the Boreas GPU backend, for UWL observations only. `INSTRUMENT' marked `n/a' (not applicable) indicates that data from these instruments predated the PSRFITS format and therefore will not be in DAP. Instruments for where there were no data found are marked with \textsuperscript{!}. \url{https://resolver.caltech.edu/CaltechETD:etd-09102008-091511}\textsuperscript{1}.}}
\begin{tabular}{p{0.4\textwidth}p{0.2\textwidth}p{0.1\textwidth}p{0.2\textwidth}}
\hline
Name & INSTRUMENT & Commissioned & Credit \\
\hline
30 MHz Polarimeter                                                                                                          & n/a$^!$              & 1990(1)        & UTAS         \\
AT Correlator                                                                                                               & n/a$^!$              & 1991(1)        & ATNF         \\
Filter Bank (FB) (\citealt{1993Natur.361..613J})                                                                            & FB\_1BIT             & 1991(5)        & ATNF         \\
S2$^*$ (\citealt{543987})                                                                                                   & n/a$^!$              & 1993(9)        & ATNF, ISTS    \\
Fast Pulsar Timing Machine (FPTM)$^1$                                                                                       & FPTM\_I, FPTM\_II, MKI & 1993(4)        & ATNF, Caltech \\
\hline
Wide Bandwidth Digital Recording (WBDR) system (\citealt{Jenet_1997})                                                       & $^!$                 & 1995(1)        &  Caltech      \\
Analogue Filter Bank (AFB) (\citealt{manchester2001parkes})                                                                 & AFB, AFB\_32\_256    & 1997(16)       & ATNF, JBO, INAF         \\
Caltech Parkes Swinburne Recorder (CPSR I) (\citealt{2000ASPC..202..283V})                                                  & CPSR                 & 1998(11)       & Caltech, ATNF, Swinburne \\
Caltech Parkes Swinburne Recorder II (CPSR2) (\citealt{2003ASPC..302...57B})                                                & CPSR2                & 2002(8)        & Caltech, ATNF, Swinburne \\
Wide Band Correlator (WBC) (\citealt{2013PASA...30...17M})                                                                  & WBCORR               & 2003(3)        & ATNF         \\
\hline
Pulsar Digital Filter Bank 1 (PDFB1) (\citealt{2013PASA...30...17M}; \citealt{2004ExA....17..269F})                         & PDFB1                & 2005(3)        & ATNF         \\
Pulsar Digital Filter Bank 2 (PDFB2) (\citealt{2013PASA...30...17M}; \citealt{2004ExA....17..269F})                         & PDFB2                & 2007(3)        & ATNF         \\
Pulsar Digital Filter Bank 3 (PDFB3) (\citealt{2013PASA...30...17M}; \citealt{2004ExA....17..269F})                         & PDFB3                & 2008(6)        & ATNF         \\
Pulsar Digital Filter Bank 4 (PDFB4) (\citealt{2013PASA...30...17M}; \citealt{2004ExA....17..269F})                         & PDFB4                & 2008(14)       & ATNF         \\
ATNF Parkes Swinburne Recorder (APSR) (\citealt{2011PASA...28....1V})                                                       & APSR                 & 2008(6)        & ATNF, Swinburne         \\
\hline
Berkeley Parkes Swinburne Recorder (BPSR, HIPSR) (\citealt{10.1111/j.1365-2966.2010.17325.x}; \citealt{Price_2016})$^*$$^*$  & HIPSR\_SRCH          & 2008(12)       & Berkeley, ATNF, Swinburne   \\
CASPER Parkes Swinburne Recorder (CASPSR) (\citealt{2011PASA...28....1V})                                                   & CASPSR               & 2010(10)       & Berkeley, ATNF, Swinburne   \\
Medusa GPU backend (\citealt{2020PASA...37...12H})                                                                          & Medusa               & 2018(7)        & Fourier Space, ATNF \\
Apollo$^+$                                                                                                                  & Apollo               & 2025(present)  & Fourier Space, ATNF         \\
Boreas GPU backend                                                                                                          & Boreas               & 2025(present)  & Fourier Space, ATNF       \\
\hline

\end{tabular}
\label{tab:dataBackend}
\end{table*}

\begin{table*}
\small
\caption{\textit{Specifications of Murriyang's principal data acquisition systems used for pulsar observations since 1990. `INSTRUMENT' refers to the value of the `INSTRUMENT' parameter key in a PSRFITS file. `Bandwidth' refers to the maximum instantaneous bandwidth of the digital backend Analogue to Digital Converters, and `Sample time' and `Resolution' are the maximum time and frequency resolution respectively. Some instruments\textsuperscript{*} have software-dependent frequency resolution. The AFB was available in various modes (all single polarisation): high-resolution, single beam (from 1997)\textsuperscript{1}; standard Multibeam Survey mode (from 1997)\textsuperscript{2}; 125kHz and 250kHz  modes for the 50cm and 70cm receivers (1997-2004)\textsuperscript{3}; wide-bandwidth, single-beam mode for the 10cm receiver (from 2005)\textsuperscript{4}. The Swinburne systems were dual-polarisation\textsuperscript{5}.}}
\begin{center}
\begin{tabular}{lllll}
\hline
INSTRUMENT & Sample time (us) & Bandwidth (MHz) & Resolution (KHz) & Bit depth \\
\hline
FB\_1BIT             &       150    &   320     & 125     & 1   \\
FPTM\_I/FPTM\_II/MKI &       2.7    &   128     & 250     & 2 \\
AFB$^1$                &       80     &   256     & 500     & 1   \\
AFB$^2$                &       125    &   288     & 3000    & 1   \\
AFB$^3$                &       125    &   32/64   & 125/250 & 1   \\
AFB$^4$                &       125    &   576/864 & 3000    & 1   \\
AFB\_32\_256$^3$       &       125    &   32      & 125     & 1   \\
CPSR                 &       0.05   &   20      & $^*$   & 1, 2, 4   \\
CPSR2$^5$              &       0.016  &   128     & $^*$   & 2   \\
WBCORR               &       7.4    &   256     & 1000    & 2 (3 levels)   \\
PDFB1                &       2      &     256   & 250     & 8   \\
PDFB2                &       2      &     256   & 250     & 8   \\
PDFB3                &       64     &     1024  & 250     & 1, 2, 4, 8, 16   \\
PDFB4                &       64     &     1024  & 250     & 1, 2, 4, 8, 16   \\
APSR$^5$               &       0.016  &     1024  & 500     & 2, 4, 8 \\
CASPSR$^5$             &       0.0025 &     400   & 800     & 2, 4, 8 \\
HIPSR\_SRCH$^5$        &       64     &     400   & 400     & 2   \\
Medusa               &       32     &     3328  & 0.061   & 2, 4, 8 \\
Apollo               &       32     &     3328  & 0.061   & 2, 4, 8 \\
Boreas               &       57     &     1250  & 0.061   & 2, 4, 8 \\
\hline
\end{tabular}
\end{center}
\label{tab:dataBackend2}
\end{table*}

\subsubsection{Scope of sky coverage} 

\begin{figure*}
\centering
\includegraphics[width=450px]{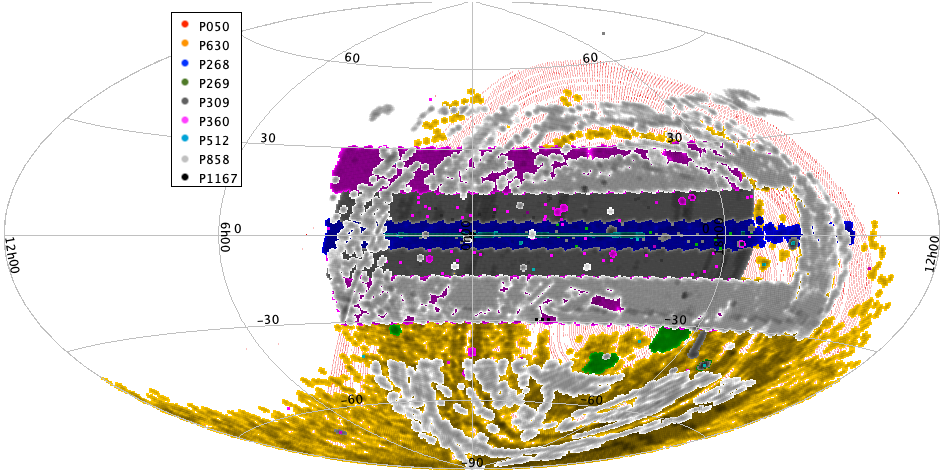}
\caption{A TOPCAT Hammer-Aitoff sky projection in Galactic coordinates of observations published in the DAP from the main pulsar surveys conducted with Murriyang over the last 30 years.}
\label{fg:Surveys}
\end{figure*}

At the time of writing, ~4.3 Petabytes of pulsar search-mode pulsar are published in the DAP, encompassing observations from both legacy and recent surveys, and pulsar and FRB follow-up campaigns. 
A total of 1235 pulsars have been discovered in surveys using Murriyang --- the first survey using the 70 cm receiver (P050) yielded a total of 298 new discoveries, and the PMPS (P268) has been the most successful to this day, finding 833 new pulsars in the galactic plane. Recent searches with new algorithms have continued to add value to these data-sets taking the total number of pulsars discovered in the PMPS to 1160, with \citealt{2025ApJ...991....6X} finding a new pulsar recently in P050 data. These examples demonstrate the importance of a comprehensive long-term data archive such as DAP.
Table~\ref{tab:dataSurveys} lists a selection of the pulsar surveys, and Figure \ref{fg:Surveys} shows the sky coverage by Project ID for the main pulsar surveys conducted over the last 34 years. 

\clearpage
\onecolumn
\begin{longtable}{p{0.05\textwidth}p{0.7\textwidth}p{0.1\textwidth}}
\caption{\textit{Known pulsar surveys and targeted searches conducted with Murriyang. `Date' refers to the range of observation dates for data in DAP. PIDs where the data in DAP are incomplete are marked with an \textsuperscript{x} --- data likely missing deemed lost or corrupt. PIDs where data are continuing to be added are marked with a \textsuperscript{+}.} PIDs with no data found to date are marked with \textsuperscript{!}, and PIDs with no data on DAP are marked with \textsuperscript{*}. PIDs marked `n/a' (not applicable) are projects that predated the PID indexing scheme.}
\label{tab:dataSurveys}
\\\hline
PID & Title & Date \\
\hline
n/a$^!$  & Large Magellanic Cloud pulsar search &                                       1980\textendash{}1981 \\
n/a$^!$  & Radio pulsars in the Magellanic clouds &                                     1980\textendash{}1982 \\
n/a$^!$  & A high-frequency survey of the southern Galactic plane for pulsars &         1988                  \\
P11$^!$   & 50cm pulsar survey &                                                         1990                  \\
P050  & 70cm pulsar survey and timing &                                             1991\textendash{}1994 \\
P151$^!$ & A pulsar search targetted towards unidentified Gamma-ray sources &           1994\textendash{}1996 \\
P153$^!$ & A Search for Pulsars from an X-Ray Sample &                                  1994                  \\
P155$^!$ & A Search for Pulsars in Supernova Remnants &                                 1994                  \\
P166$^!$ & A Directed High Frequency Search for Pulsars near the Galactic Centre &      1995                  \\
P168$^!$ & A Search for Pulsars in Supernova Remnants &                                 1995                  \\
P169$^!$ & Are there any pulsars in the Galactic Centre? &                              1995                  \\
P247$^!$ & A deep survey for pulsars in the Magellanic Clouds &                         1996                  \\
P256$^!$ & Searching for pulsars in the Carina and Crux spiral arms &                   1997\textendash{}1998 \\
P263$^!$ & Baseband searching for ultrafast pulsars &                                   1997\textendash{}2003 \\
P267$^!$ & Search for pulsar in SN 1987A &                                              1997\textendash{}1999 \\
P268 & Pulsar multibeam survey &                                                    1997\textendash{}2004 \\
P269 & A deep pulsar survey of the Small Magellanic Cloud &                         1997\textendash{}2001 \\
P272$^!$ & High sensitivity pulsar search of 47 Tuc at 21cm &                           1997         \\
P273$^!$ & A search for pulsar counterparts to EGRET sources &                          1997                  \\
P275$^!$ & A search for radio pulsars in EGRET sources &                                1998                  \\
P279$^!$ & Search for relativistic binary pulsars in globular clusters &                1998                  \\
P281$^!$ & New fast pulsars in SNRs RCW 103 and G11.2-0.3 &                             1998                  \\
P282 & Timing of millisecond pulsars in 47 Tucanae &                                1998\textendash{}2013 \\
P296$^!$ & A search for submillisecond pulsars in Globular Clusters &                   1998                  \\
P302$^!$ & Improving the scintillation constraints on PSR1706-44 &                      1998                  \\
P303 & Search and Timing of Pulsars in Globular Clusters &                          1998\textendash{}2002 \\
P305$^!$ & A search for a young pulsar in SGR G0.9+0.1 &                                1998                  \\
P309 & An intermediate-latitude millisecond pulsar survey &                         1998\textendash{}1999 \\
P322$^!$ & Search for radio pulsars toward selected GEV sources	 &                      1999                  \\
P326$^!$ & Search for radio pulsations from A0538-66 &                                  1999                  \\
P327$^!$ & A search for young pulsars in three composite SNRs &                         1999\textendash{}2000 \\
P328$^!$ & Deep search for radio pulsations from Magnetars &                            1999                  \\
P331$^!$ & Search for pulsars in the NVSS catalogue &                                   1999                  \\
P332$^!$ & Search for pulsed radio emission from anomalous X-ray pulsars &              1999                  \\
P346$^!$ & Search for Pulsations from a new 87 ms SMC X-ray Pulsar &                    2000                  \\
P358$^!$ & Deep pulse searches of three pulsar wind nebula candidates &                 2001                  \\
P359$^!$ & A search for radio pulsations from isolated neutron stars &                  2001                  \\
P360 & A high-latitude millisecond pulsar survey &                                  2001\textendash{}2002 \\
P365 & A search for giant pulses (in PSR J0537-6910) &                              2001\textendash{}2003 \\
P366 & Parkes multibeam high-latitude pulsar survey &                               2001\textendash{}2003 \\
P385$^!$ & Searching for the Pulsar in SNR G292.0+1.8 &                                 2001                  \\
P390$^!$ & Searching for a Pulsar in SNR G16.73+0.08 &                                  2001                  \\
P394$^!$ & A search for radio emission from magnetars &                                 2002                  \\
P396$^+$ & Deep searches for young and `radio-quiet' pulsars &                          2002\textendash{}2005 \\
P397$^!$ & Searching for pulsars in new supernova remnants &                            2002                  \\
P401$^!$ & A pulsar search for four Egret sources in the Galactic halo &                2002                  \\
P403$^!$ & A survey for pulsars at very high galactic latitudes &                       2002                  \\                      
P406$^!$ & A search for pulsars in mid-latitude EGRET error boxes &                     2002\textendash{}2004 \\
P413$^!$ & A possible sub-millisecond gamma-ray pulsar &                                2002                  \\
\caption{\textit{Continued...}}
\\\hline
PID & Title & Date \\
\hline
P427$^+$ & Timing and searching millisecond pulsars in Globular Clusters &              2003\textendash{}2015 \\
P434$^!$ & A deep search for isolated millisecond pulsars in NGC 6266 &                 2003                  \\
P441$^!$ & Searching for pulsars in SNRs N206 and B0453-685 &                           2003                  \\
P442$^!$ & Searching for Globular Cluster MSPs at 50cm &                                2003                  \\
P448$^+$ & Search for radio pulsations from four X-ray msec pulsars &                   2003                  \\
P461 & A pilot 50 cm pulsar survey &                                                2004\textendash{}2005 \\
P464$^!$ & Search for old pulsars associated with EGRET unidentified sources &          2004                  \\
P471$^+$ & Deep multibeam pulsar survey at 50 < l < 60 &                                2004\textendash{}2006 \\
P473$^!$ & A search for double-pulsar binaries &                                        2004                  \\
P477$^+$ & The Perseus Arm pulsar multibeam survey &                                    2004\textendash{}2006 \\
P487$^!$ & Deep searches for young pulsars in `shell' supernova remnants &              2004\textendash{}2005 \\            
P490$^!$ & A search for extra-galactic giant pulses &                                   2005                  \\
P491$^+$ & Pulsar searches of the Galactic Centre &                                     2005                  \\
P505 & Pulsar survey of the Canis Major dwarf Galaxy &                              2005\textendash{}2006 \\
P512$^x$ & A methanol multibeam pulsar survey &                                         2006\textendash{}2008 \\
P535$^*$ & A Census of Pulsar Emission &                                                2006\textendash{}2008 \\
P589$^!$ & Searching for pulsars associated to recently detected Pulsar Wind Nebulae and X-ray point sources in Supernova Remnants & 2007 \\
P630$^+$ & The High Time Resolution Universe &                                          2008\textendash{}2014 \\
P661 & The Search for and Confirmation of Nearby RRAT Candidates &                  2008\textendash{}2011 \\                     
P675$^+$ & Radio search for gamma-ray pulsar counterparts &                             2009\textendash{}2011 \\
P680$^!$ & A search for the youngest pulsar in the Galaxy &                             2009                  \\
P682 & A high time resolution survey of the Small Magellanic Cloud &                2009                  \\
P778 & A new deep search for millisecond pulsars in globular clusters &             2010\textendash{}2014 \\
P809 & Parkes search for radio pulsars from unidentified Fermi Sources at 50 cm  &  2011\textendash{}2012 \\
P814 & Millisecond pulsar searches in unidentified Fermi sources at high Galactic latitudes & 2012\textendash{}2017 \\
P834 & Searching for the pulsar in SN1987A &                                        2013\textendash{}2019 \\
P855 & A Parkes transit survey for pulsed radio emission during windstows and maintenance & 2013\textendash{}2015 \\
P858$^+$ & SUPERB - A SUrvey for Pulsars \& Extragalactic Radio Bursts &                 2014\textendash{}2015 \\
P859 & Searching towards the Galactic Centre region for pulsed radio emission &     2014                  \\
P864 & A Search for the Intergalactic Magnetic Field &                              2014                  \\
P865 & Commensal searches for microhertz gravitational waves and fast radio bursts: A pilot study & 2014  \\
P867 & Searching for Radio Millisecond Pulsars in a New Set of Fermi Sources &      2014                  \\
P869 & Ultra-Deep searches for pulsars around Sgr A* &                              2014                  \\
P883 & The Parkes All Radio Transients in the skY (PARTY) survey &                  2015                  \\
P886 & Searching for pulsars using the CSIRO Pulsar Data Archive &                  2015                  \\
P892$^+$ & SUPERBx - The SUrvey for Pulsars \& Extragalactic Radio Bursts Extension &    2015\textendash{}2019 \\
P895 & Where are the gravitational waves? &                                         2015\textendash{}2020 \\
P914 & Searching for pulsars from steep-spectrum MWA candidates &                   2016\textendash{}2017 \\
P942 & Searching for giant pulses from pulsars in the Large Magellanic Cloud &      2017                  \\
P970 & Searching for millisecond pulsars towards unidentified Fermi sources using the Ultra-wideband receiver & 2018 \\
P982 & The first coherently de-dispersed search for new pulsars in Southern globular clusters & 2018 \\
P986 & Targeted search of steep spectrum sources with the Ultra-Wideband receiver & 2018\textendash{}2019 \\
P991 & A Pulsar Survey Towards the Galactic Centre With the Ultra-Wideband Low Receiver & 2018\textendash{}2019 \\
P998 & Terzan 6 - The Next Terzan 5? &                                              2018                  \\
P1000 & A deep search for radio pulsars around hot subdwarf stars in compact binaries & 2018 \\
P1006 & The first ultra-high time resolution coherently de-dispersed search for new pulsars in globular clusters & 2019 \\
& & \\
\caption{\textit{Continued...}}
\\\hline
PID & Title & Date \\
\hline
P1022 & The first well-calibrated, coherent de-dispersed search for pulsars in 47 Tucanae & 2019\textendash{}2020 \\
P1034 & Follow-up of the first millisecond pulsar discoveries in Omega Centauri and a new coherently de-dispersed survey & 2019\textendash{}2020 \\
P1052$^*$ & Pulsar Radio Emission Statistics Survey (PRESS) &                           2020\textendash{}2021 \\
P1067 & A wide-band survey of Baade's Window in the time and frequency domains &    2020\textendash{}2021 \\
P1082 & Revealing Mercer 5's Pulsar Population with Parkes &                        2020\textendash{}2021 \\
P1087 & Vela in the LMC: A Search for the Pulsar in SNR B0453-685 &                 2020\textendash{}2021 \\
P1090 & Searching for millisecond pulsars around blazar OJ 287 &                    2020\textendash{}2021 \\
P1156 & Targeted Search of Fast Radio Bursts from Globular Clusters in Centaurus A & 2022 \\
P1163 & Searching for millisecond pulsars in Sagittarius dwarf spheroidal (Sgr dSph) galaxy using the Ultra-wideband Low receiver & 2022\textendash{}2023 \\
P1167 & A wide-band survey of the Galactic Centre in the time and frequency domains & 2022\textendash{}2023 \\
P1193 & Searching for pulsars among Fermi unassociated LAT sources with RACS &      2023 \\
P1197 & Searching for pulsars in four supernova remnant &                           2023 \\
P1211 & A targeted search for Millisecond Pulsars in Galactic Plane towards steep spectrum radio sources & 2023\textendash{}2024 \\
P1226 & A targeted search for Fast Radio Bursts in dwarf satellite galaxies &       2023\textendash{}2024 \\
P1238 & Searching for pulsars in new Galactic pulsar wind nebula candidates &       2023\textendash{}2025 \\
P1330 & Determining Pulsar NGC6316A's Orbit and the Search for Additional Pulsars & 2024                  \\
P1336 & A Search for Hidden Pulsars from the MeerKAT Bulge Survey &                 2024\textendash{}2025 \\
P1347 & A Targeted Pulsar Search in the Galactic Center and Bulge &                 2024\textendash{}2025 \\
\hline

\end{longtable}

\clearpage

\twocolumn

\subsubsection{Collections containing historically important discoveries}

Within the daily stream of data from observations at Murriyang, the DAP contains archival data that lead to several notable discoveries --- a selection of those are shown in Table \ref{tab:dataImportant}. The data file containing the discovery, the DOI of the collection the files are in, and the publication reference are all listed. 

Access to the archive is continuing to grow, and this is reflected in the NASA ADS\footnote{\url{https://ui.adsabs.harvard.edu}} publication statistics - 15 peer-reviewed papers acknowledged the use of DAP data in their work in 2024.

\begin{table*}
\small
\caption{\textit{DAP collections containing important discoveries.}}
\begin{center}
\begin{tabular}
{p{0.35\textwidth}p{0.2\textwidth}p{0.35\textwidth}}
\hline
Description & Discovery File & DAP Collection DOI  \\
\hline
Closest known MSP (J0437-4715) (\citealt{1993Natur.361..613J})              & S13804\_7.sf           & https://doi.org/10.4225/08/57188974623F6 \\
Pulsar A of the double pulsar (J0737-3039A) (\citealt{2003Natur.426..531B}) & PH0042\_004B1.sf       & https://doi.org/10.4225/08/598c2d9103f0c \\
The Lorimer Burst (first FRB) (\citealt{2007Sci...318..777L})               & SMC021\_00861.sf        & https://doi.org/10.4225/08/5819628e4fed9 \\
First MSP in the Galactic Centre (\citealt{2024ApJ...967L..16L})            & uwl\_240325\_000341.sf & https://doi.org/10.25919/fpnn-4r75       \\
\hline

\end{tabular}
\end{center}
\label{tab:dataImportant}
\end{table*}

\subsection{Additions to the archive}

Some archival Murriyang pulsar data-sets are gradually being added to DAP. These tend to include very large surveys (for example  the `High Time Resolution Universe Pulsar Survey' (P630, \citealt{2010MNRAS.409..619K}) --- there is an ongoing effort to collate these `missing' data, and archive them as resources become available --- currently, over 25 Terabytes of data from this survey are being added to the archive annually.

\section{Data Dissemination}

Providing data accessibility through requests based on a search and filter methodology is a core part of the archive. The DAP provides both a simple and advanced search interface, with the latter allowing filtering on source name, position, observation frequency or receiver.

DAP is also Virtual Observatory\footnote{\url{https://ivoa.net}} (VO) compatible. VO tools such as \textsc{TOPCAT}\footnote{\url{https://www.star.bris.ac.uk/~mbt/topcat/}} or \textsc{PyVO}\footnote{\url{http://github.com/astropy/pyvo}} can be used to query pulsar observations using the Table Access Protocol (TAP) and Simple Cone Search (SCS), and cross-match sources with other catalogues --- for example, Figure \ref{fg:Surveys} was generated by a TAP query in \textsc{TOPCAT}. A configured query can also be used to generate data download links.

After selecting the required data, a user can choose a download method. In 2023, DAP moved to object storage --- this resulted in considerable improvement in the access of large Terabyte-scale collections. The object store supports transfer protocols such as the AWS Command-Line Interface\footnote{\url{https://aws.amazon.com/cli}}, Rclone\footnote{\url{https://rclone.org}} and Globus\footnote{\url{https://www.globus.org}}.

\section{Data Analysis and Visualisation}

There are a number of ways a user can interact with data from the DAP, and this will depend on the nature of the research or required scientific outcome. 

Re-analysing search-mode data from all-sky surveys can lead to new pulsar discoveries, or allow studies of single-pulse phenomenology of known sources. Fold-mode data can be used for a variety of astrophysical studies, such as pulsar timing, profile evolution, polarimetry, and the properties of the Inter-Stellar Medium (ISM), and the use cases will depend on the required scientific outcome.

A PI or other astronomer may wish to use conventional command-line pulsar data analysis packages, for example, \textsc{PSRCHIVE}\footnote{\url{https://psrchive.sourceforge.net}}(written in C++ with Python wrappers) for folding/calibration or \textsc{PRESTO}\footnote{\url{https://github.com/scottransom/presto}} (in C with Python wrappers for some routines) for pulsar searching --- these packages all read files in the PSRFITS format. The user may also wish to pre-configure packages to suit their system, and implement their own algorithms.
 
\subsection{Introducing the \textsc{PFITS} package}

Many updates to the conventional pulsar data analysis packages mentioned above have been conducted since they were published, in order to work with wide-band data such as those from the UWL receiver, although they each work with data from specific observing modes (fold- or search-mode). However, here we introduce the \textsc{PFITS}\footnote{\url{https://github.com/too043/pfits.git}} package --- written in C, it is an alternative to conventional tools and provides routines and utilities for working with PSRFITS format files from all the different observing modes. Some examples of commonly-used routines are:
\begin{itemize}
  \item\textbf{pfits\_describe} --- prints the header information
  \item\textbf{pfits\_fv} --- interrogates the file metadata interactively
  \item\textbf{pfits\_plot} --- interactively plots the astronomy data
  \item\textbf{pfits\_zapProfile} --- interactively removes interference
  \item\textbf{pfits\_frb} --- interactively plots FRB candidates in a pulsar search-mode file
\end{itemize}

\begin{figure*}
\centering
\includegraphics[width=400px]{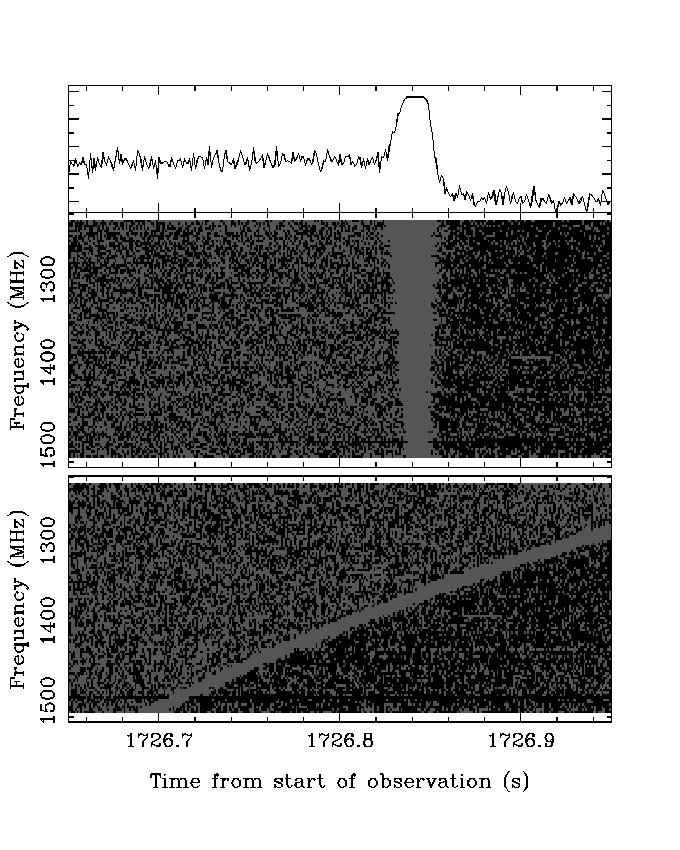}
\caption{The output of pfits\_frb: the profile of the Lorimer Burst is shown in the top plot, de-dispersed in the centre and dispersed at the bottom. The profile shows significant clipping in this beam (the discovery beam) of the Multibeam receiver because the pulse saturated the available dynamic range.}
\label{fg:pfits_frb}
\end{figure*}

Figure \ref{fg:pfits_frb} demonstrates output from the pfits\_frb routine --- a user can zoom in time to display a window around the dispersed pulse of an FRB, and the pulse is then de-dispersed on the fly.

\section{Pulsar Data Archiving Challenges and Future Requirements}

Pulsar data archiving into the future provides the following challenges, including handling increasingly large data volumes, the importance of provenance for reproducibility, the hunt for missing data, and Cloud-based archiving as a means to provide global access to data products.

\subsection{Managing high data volumes}

The Cryogenically-cooled Phased Array Feed is the next generation receiver for Murriyang --- consisting of 72 beams and designed for large-scale surveys, capable of recording an instantaneous bandwidth in two bands from 700-1200 and 1100-1950 MHz, with expected data rates of up to ~80TB per hour depending on the observation mode and configuration. Archiving of these data will present considerable challenges, and development is underway to enable near real-time ingest of high-volume data into DAP.

\subsection{Cloud-based pulsar data processing}

Cloud-based storage and compute platforms such as Amazon Web Services (AWS), are experts in handling large data volumes. These services could be used to provide a mirror of DAP data, allowing easily configurable global access to the archive, and scaleable compute infrastructure for data processing, encouraging a `User to the data' model. We are currently trialling this model in CSIRO's Earth Analytics Science and Innovation platform (EASI\footnote{\url{https://research.csiro.au/easi}}), which runs on AWS infrastructure, and is accessible to anyone who applies for and is granted an EASI account.

\subsection{Archival data recovery}

The DAP archive is by no means a complete set of collections, and in fact data are recovered from tape as and when they are discovered. The data are checked, sorted and converted to PSRFITS format if required prior to adding to an existing DAP collection, or forming a new collection. We are currently recovering archival data and publishing these in DAP at a rate of approximately 30TB per year.

A proportion of Parkes pulsar data are not available in the archive. Of the ~400 project IDs to date, ~100 are currently missing. These data are likely found in University cupboards, simply lost, or deemed junk data.

\section{Conclusion}
Murriyang remains an instrument at the cutting edge of the field of radio astronomy. This is due mainly to investment in its receiver and digital acquisition systems over the years, but also to the availability of its data products in long-term archives such as DAP.

In this paper we have provided an update on the archive status, and demonstrated the importance of storing large volumes of pulsar data for re-processing by modern algorithms resulting in new discoveries. We introduced the \textsc{PFITS} package for processing of PSRFITS format data, and touched on the future use of Cloud platforms for scaleable processing workflows without the need to move vast volumes of data.

\section{Acknowledgements}

The Parkes radio telescope is part of the Australia Telescope National Facility (grid.421683.a) which is funded by the Australian Government for operation as a National Facility managed by CSIRO.
This paper includes archived data obtained through the Parkes Pulsar Data archive on the CSIRO Data Access Portal (http://data.csiro.au).
We acknowledge the Wiradjuri people as the Traditional Owners of the Observatory site.

\bibliography{refs}

\clearpage

\end{document}